\documentclass[twocolumn,showpacs,preprintnumbers,amsmath,amssymb]{revtex4}
\usepackage{graphicx}
\usepackage{dcolumn}
\usepackage{bm}
\begin{document}
\title{Searching for statistical equilibrium in a dynamical multifragmentation path}
\author{A. H. Raduta$^{a,c}$, M. Colonna$^{a,b}$, M. Di Toro$^{a,b}$}
\affiliation{
        $^a$LNS-INFN, via S.Sofia 62, I-95123, Catania, Italy\\
	$^b$Physics and Astronomy Dept. University of Catania, Italy\\
        $^c$NIPNE, RO-76900 Bucharest, Romania}

\begin{abstract}
A method for identifying statistical equilibrium stages in dynamical multifragmentation
paths as provided by transport models, already successfully tested for 
the reaction $^{129}$Xe+$^{119}$Sn at 32 MeV/u 
is  applied here to a higher energy reaction, $^{129}$Xe+$^{119}$Sn at 50 MeV/u.
The method evaluates equilibrium from the point of view of the microcanonical multifragmentation model (MMM)
and reactions are simulated by means of the stochastic mean field model (SMF). 
A unique solution, corresponding to the maximum population of the system phase space, was identified 
suggesting that a huge part of the available phase space is occupied even in the case
of the 50 MeV/u reaction, in presence of a considerable amount of radial collective flow. 
The specific equilibration time and volume
are identified and differences between the two systems are discussed.
\end{abstract}
\maketitle

\section{Introduction}

From more than 20 years, {\em statistical equilibrium} was a basic ingredient in many theoretical \cite{bauer,Gross1,Bondorf,Randrup,mmm} and experimental nuclear multifragmentation studies. 
This was motivated by the good agreements obtained over the time between various observables related to the asymptotically resulted fragments and various statistical multifragmentation models. 

However, the agreement in the asymptotic stage between experiments and statistical models 
predictions is only a {\em necessary} condition for equilibrium. For fully ``proving'' equilibrium
one has to have direct access to the {\em primary decay} stage of the reaction.
 In fact, different primary configurations could lead to the same final results,
because of compensative effects between primary and secondary emission mechanisms.
Experimentally it is difficult to ``measure'' the primary decay stage: this signal is distorted by the 
secondary particle emission. An elegant solution is to perform a statistical analysis on the primary decay stage 
as predicted by a dynamical model yielding results in good agreement with experimental data. 
One of such models is the Stochastic Mean Field (SMF)\cite{mac,frankland,buu}. 

In Ref. \cite{dyn-1} a statistical analysis was performed on the dynamical path provided by SMF with the microcanonical multifragmentation model (MMM) \cite{mmm}. 
There, an effective method for identifying the 
{\em equilibrated stage} in the dynamical path was proposed and  successfully tested for the $^{129}$Xe+$^{119}$Sn 
at 32 MeV/u reaction. A fully equilibrated source was identified at 140 fm/c, with a corresponding volume of
3.4 {\em $V_0$} (where $V_0$ is the volume of the source at normal nuclear density).
Herein we apply the same analysis to the higher energy reaction $^{129}$Xe+$^{119}$Sn at 50 MeV/u as 
described by the SMF model. 
In this way one can discuss the possible occurrence of statistical equilibrium 
even for a more explosive system, having a considerable amount of
radial collective flow. 
Important differences between the two reactions are supposed to appear. Since at 50 MeV/u
the emitted fragments are generally smaller, one can find a large variety of 
source configurations that can fit the fragment partitions. However, we will show 
that the analysis of the corresponding fragment kinetic properties allows to 
clearly identify a unique solution.

\section{Brief review of the employed models}

Let us first give a brief description of the two models involved in this analysis.
According to the SMF theory \cite{mac}, 
 during the expansion phase that follows the initial collisional shock 
the system encounters volume (spinodal) and/or surface instabilities, that
lead to its break-up into many pieces. 
As in mean-field approaches, the system is represented by the one-body distribution function
in phase space, that evolves according to the self-consistent nuclear (+ Coulomb) potential
and to the stochastic collision integral, that accounts for the residual two-body interaction.  
In the approach considered, fluctuations are introduced in an approximate way \cite{mac}.
Their amplitude  is essentially 
determined by the degree of thermal agitation present in the system. 
Then fluctuations are amplified by the unstable mean-field, leading to
the formation of fragments, whose properties reflect the structure of 
the most unstable collective modes.
Several multipoles are excited with close probabilities, leading to a large variety
of fragment configurations.\cite{rep}. 

Concerning more technical aspects, such as fragment recognition, we follow
a coalescence procedure of the one-body density. 
Fragment excitation energies are calculated adopting the local density 
approximation,  
by subtracting 
the Fermi motion (associated with the local density)
from the fragment kinetic energy (taken in the fragment reference frame) 
\cite{rep,AlfioBOB}.


The MMM model \cite{mmm} describes the break-up of a statistically equilibrated nuclear source
defined by the parameters: mass number ($A$), atomic number ($Z$), excitation energy ($E$) and freeze-out volume ($V$). 
The model assumes equal probability between all configurations $C:\{A_i,Z_i,\epsilon_i, 
{\bf r}_i,{\bf p}_i,~~i=1,\dots,N\}$ (the mass number, the atomic number, the excitation energy, the position
and the momentum of each fragment $i$ of the configuration $C$, composed of $N$ fragments) subject
to microcanonical constraints: $\sum_i A_i=A$, $\sum_i Z_i=Z$, $\sum_i {\bf p}_i=0$, $\sum_i {\bf 
r}_i\times{\bf p}_i=0$, $E$ - constant. The level density of a given fragment (entering the statistical weight of a 
configuration) is taken to be of Fermi-gas type adjusted with the cut-off factor $\exp(-\epsilon/\tau)$: 
$\rho(\epsilon)=\rho_0(\epsilon) \exp(-\epsilon/\tau)$ \cite{prc2002} that counts for the dramatic decrease of the lifetime of fragment excited states respective to the freeze-out specific time as the excitation energy increases. MMM can work within two freeze-out hypotheses: (1) fragments are treated as hard spheres placed into a spherical freeze-out recipient and are not allowed to overlap each-other or the recipient wall; (2) fragments may be deformed and a corresponding free-volume expression is approaching the integration over fragment positions \cite{prc2002}. 
Hypothesis (2) is more adequate for the present study since it allows exploration of higher density configurations. 
Indeed, our dynamical studies seem to indicate that a good amount of
statistical equilibrium is reached at a pre-fragment level, i.e. when the produced
fragments are not fully separated yet \cite{dyn-1}. This condition cannot be always matched by the space 
configurations related to the hypothesis (1), which presents severe constraints for
low freeze-out volumes. 
Moreover, 
as resulting from \cite{prc2002}, at relatively small freeze-out densities the results of the two approaches roughly coincide. This is why we perfom the present investigation within hypothesis (2). The model employs a Metropolis-type simulation for estimating the average value of any system observable (see Refs. \cite{mmm} for more details). While MMM includes a secondary decay stage, this stage is not necessary for the present analysis because we aim at a direct investigation of the (primary) {\em break-up} stage. A feature of this model, particularly important for the present study is the possibility of including {\em flow} in the primary decay stage. 
We include a standard flow velocity profile $v=v_0 (r/R)^{\alpha}$ \cite{prc2002}, 
where $r$ is the distance of a given fragment from the system c.m. 
and $R$ is the freeze-out recipient radius. In MMM the total flow energy is {\em microcanonically} conserved. Therefore, $v_0$ is evaluted for each fragmentation event from the condition of conservation of the system flow energy. The flow exponent $\alpha$ will be varied between the values 1 and 2. 

\section{Statistical analysis}

Using MMM we perform the analysis proposed in Ref. \cite{dyn-1} in order to identify a possible statistical equilibration stage in the dynamical path of the reaction $^{129}$Xe+$^{119}$Sn at 50 MeV/u as provided by SMF.
Within SMF the {\em dynamical freeze-out} time is defined as the time when the fragment formation process is over
and therefore (intermediate mass) fragment multiplicities do not evolve any longer. This time was identified to be around 200 fm/c. 

In order to wash-up pre-equilibrium effects, which should appear in the dynamical simulation, 
only intermediate mass fragments (IMF) (i.e. fragments with $Z\ge3$) are considered from both SMF and MMM simulations. Due to the large Coulomb repulsion  among primary fragments 
and the possible presence of a collective flow, 
it is reasonable to assume that
from the {\em equilibrated freeze-out}, i.e. the time when the system could have reached full phase space occupation, 
to the readily identified {\em dynamical freeze-out} the major difference
will be the volume in which fragments are located. 
In other words we consider that all source variables except {\em volume} are roughly preserved.
The fragment properties included in our analysis are connected to
relevant 
inclusive observables that can be predicted by both models: fragment multiplicity
and distributions and fragment internal excitation energy. 
We will fit these properties, but {\em not} the volume. 
For finding the best fit we have to minimize the following error function \cite{dyn-1}: 
\begin{eqnarray}
{\cal E}_1&=&\{3\left[f(\left<A_{bound}\right>)+f(\left<Z_{bound}\right>)\right] \nonumber \\
        &+&\left[\sum_{N_{IMF}}f[\left<{\rm d}N/{\rm d}N_{IMF}\right>]/\sum_{N_{IMF}}1\right] \nonumber \\
	&+&f(\left<\epsilon_{IMF}\right>) 
        +\sum_{i=1}^3 f(\left<Z_{{\max}i}\right>)/3\}/9
\end{eqnarray}
where $\left< \cdot \right>$ stands for average, $A_{bound}$ and $Z_{bound}$ are the bound mass and 
charge (sum of the mass number and, respectively, atomic number of all IMF's from a given event), $N_{IMF}$ is
the number of IMF's, $\epsilon_{IMF}$ is the fragment excitation energy per nucleon and $Z_{{\max}i}$ with $i=1,2,3$ are the largest, second largest and third largest charge from one fragmentation event. Further, 
$f(x)=\left|2(x_s-x_d)/(x_s+x_d)\right|$, where the indexes $s$ and $d$ stand for ``statistic'' and
``dynamic''.
\begin{figure}
\includegraphics[height=9cm,angle=270]{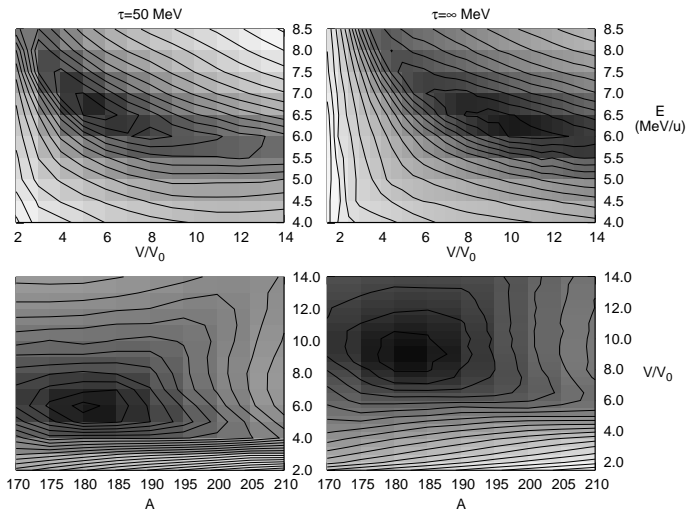}
\caption{Contour plots of the error function ${\cal E}_1$ [see eq. (1)] for $\tau=50$ MeV (left column) and $\tau=\infty$ MeV (right column). Cuts in the $(V/V_0,E)$ (upper panel) and in the $(A,V/V_0)$ (lower panel) planes corresponding to the minimum ${\cal E}_1$ coordinates. Darker regions correspond to smaller ${\cal E}_1$; units are relative. \\[-.7cm]}
\end{figure}

For finding the coordinates of the minimum of ${\cal E}_1$ we variate the MMM parameters $A$, $E$, $V$ and $\tau$ in wide 
ranges thus constructing a four-dimensional grid. The ranges are $A:\left[170,210\right]$, $E:\left[4,8.5\right]$ MeV/u, $V/V_0:\left[2,14\right]$, $\tau=20, 50, 100, {\infty}$ MeV. The source is considered to have the $A/Z$ ratio of the $^{129}$Xe+$^{119}$Sn reaction. 
Like for the case of the 32 MeV/u incident energy for a given value of $\tau$ one gets a unique minimum
for ${\cal E}_1$. This is illustrated in Fig. 1 where cuts in ${\cal E}_1$ corresponding to $A_{min}$ and $E_{\min}$ are represented for two values of $\tau$: 50, $\infty$ MeV (where the index $'{\min}'$ refers to the coordinates of the minimum value of ${\cal E}_1$). 
However, apart from the $^{129}$Xe+$^{119}$Sn at 32 MeV/u reaction, in the present case one cannot get an 
unambiguous source identification only by minimizing ${\cal E}_1$. Indeed, 
it is possible to find a solution for any value of the considered $\tau$ and,
as resulting from Fig. 2, the curve representing the minimum value of 
${\cal E}_1$ versus $\tau$ is rather flat. In other words, primary fragment distributions and their internal 
excitation energies are not sufficient for proving the occurrence of equilibrium. 

So, in order to better identify  the {\em equilibrated source} we have to analyze the reaction kinematics too. More 
specifically, we fit the fragment kinetic distributions and the positions of the three highest charged fragments corresponding to the {\em dynamical freeze-out}. 
The error function employed in this fit is:
\begin{eqnarray}
{\cal E}_2=\left[\sum_{z=5}^{40} f\left(\left< k(z)\right>\right)/36 
          +\sum_{i=1}^3 f(\left< r\left(Z_{{\max}i})\right>\right)/3\right]/2
\end{eqnarray}
($k$ denotes the fragment kinetic energy and $r$ denotes its position; in practice, for increasing the accuracy of the ${\cal E}_2$ estimator, a smoothed $k(z)$ curve was used and the z interval was spanned with a step of 5 units). 
But how can one get ``statistical'' information about the {\em dynamical freeze-out} in order to perform this fit?  One simply has to propagate the primary fragments in their mutual Coulomb field, taking into account also the
presence of collective flow effects, 
from their freeze-out positions as generated by MMM up to $\widetilde{V}_{IMF}=16.73~ V_0$ which is the volume corresponding to the {\em dynamical freeze-out}. 
(Herein we use the notations from Ref. \cite{dyn-1}: $V$ is the volume of the smallest sphere which {\em totally}
includes all fragments; $\widetilde{V}$ the volume of the smallest sphere which {\em totally} 
includes all fragments {\em and} has the center located in the center of mass of the system).
Such propagations were performed for sources corresponding to the minimum ${\cal E}_1$ coordinates for each 
of the considered values of $\tau$. The flow energy, $E_{fl} = 
\sum_i{1\over 2}m_i v_i^2$ ($m_i$ and $v_i$ being respectively the mass and the flow energy of the fragment $i$), 
and the flow exponent were varied in the ranges: $E_{fl}: [0,1.6]$ MeV/u and $\alpha_{fl}: [1,2]$. An unique minimum of the ${\cal E}_2$ function was found for each of the considered
values of $\tau$. One can observe this behavior in Fig. 3 for the case $\tau={\infty}$. 
The obtained dependence of the ${\cal E}_2$ minima of $\tau$ is given in Fig. 2. This time we observe a monotonic
dependence of ${{\cal E}_2}_{\min}$ of $\tau$. The global error function ${\cal E}={\cal E}_1+{\cal E}_2$ exhibits 
a well defined minimum as well (from now on we denote by ${\cal E}$ this {\em minimum} value). This minimum corresponds to $\tau=\infty$.

It should be noticed that the full set of inclusive observables considered in the total
estimator, ${\cal E}$,  is quite exhaustive. Most of them coincide with the observables usually exploited in the comparison with experimental data.
 
For the sake of completness, we test the ${\cal E}_2$ estimator in the case of the 32 MeV/u reaction, which was already analyzed in Ref. \cite{dyn-1}. There it was shown that fragment kinetic energies and positions were compatible with the source identified only with ${\cal E}_1$ for a zero flow energy.
In agreement with Ref. \cite{dyn-1} results, the minimum of the ${\cal E}_2$ versus $\tau$ 
curve sharply points the value $\tau=\infty$, as ${\cal E}_1$ does, see Fig.4. 
In the ($E_{fl}$, $\alpha_{fl}$), $\tau=\infty$ plane ${\cal E}_2$ minimizes inside the $E_{fl}$=0 MeV region, as shown in Fig.5, which confirms the findings of Ref. \cite{dyn-1}. 
The sharper minimum of the 
${\cal E}={\cal E}_1+{\cal E}_2$ estimator observed for the 32 MeV/u reaction compared to the 
50 MeV/u situation indicates a more advanced degree of statistical equilibrium reached in  
the 32 MeV/u reaction.

Finally, it is worth noticing that since hypothesis 1) of the MMM model induces artificial geometrical 
constraints at small volumes, it is easy to infer that, for example, in the case of 32 MeV/u one 
would not find any physically meaningfull minimum (i.e deep enough to insure a good reproduction of
the data under study). On the other hand, for the 50 MeV/u reaction, the system being {\em less} affected by these 
constraints, the result would be closer to the evaluations obtained within hypothesis 2). 

\begin{figure}
\includegraphics[height=6cm]{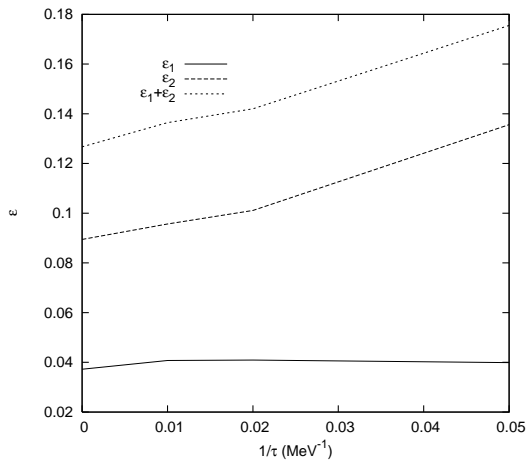}
\caption{The error functions ${\cal E}_1$, ${\cal E}_2$ and ${\cal E}$=${\cal E}_1$+${\cal E}_2$ versus $1/\tau$.\\[-.7cm]}
\end{figure}
\begin{figure}
\includegraphics[height=6cm, angle=0]{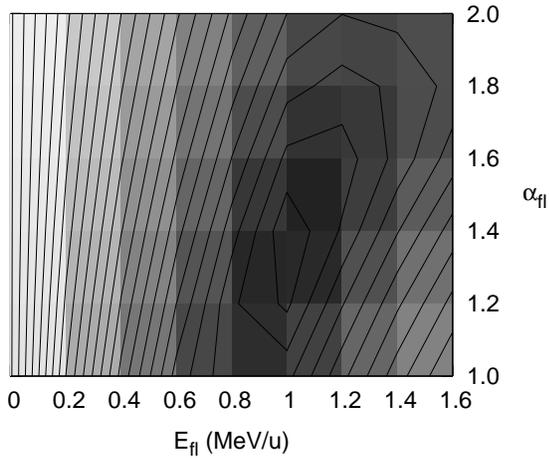}
\caption{Contour plot of the function ${\cal E}_2$ in the plane $(E_{fl},\alpha_{fl})$ corresponding to $\tau=\infty$. Units are relative; darker regions correspond to smaller values of ${\cal E}_2$. \\[-.7cm]}
\end{figure}

\begin{figure}
\includegraphics[height=6cm]{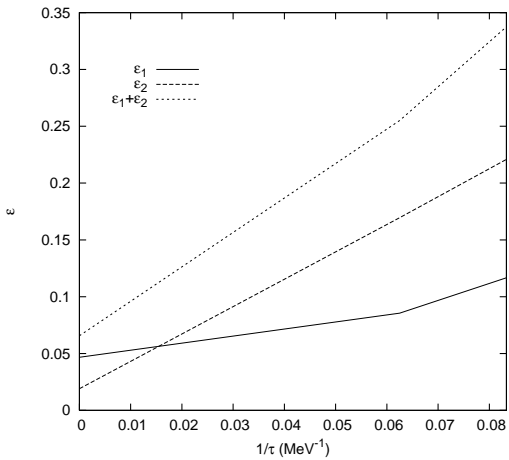}
\caption{The error functions ${\cal E}_1$, ${\cal E}_2$ and ${\cal E}$=${\cal E}_1$+${\cal E}_2$ versus $1/\tau$ for the case of the reaction Xe+Sn at 32 MeV/u.\\[-.7cm]}
\end{figure}

\begin{figure}
\includegraphics[height=6cm, angle=0]{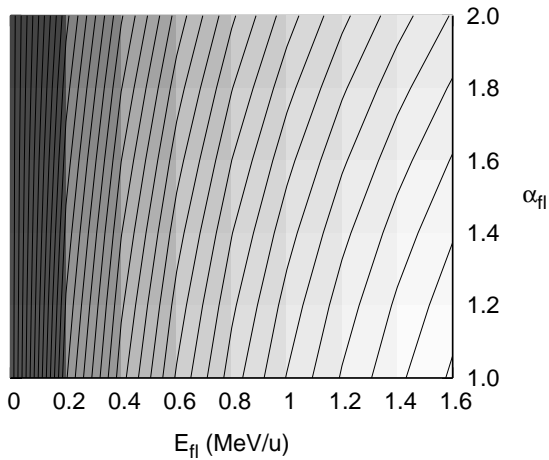}
\caption{Contour plot of the function ${\cal E}_2$ in the plane $(E_{fl},\alpha_{fl})$ for the case of the reaction Xe+Sn at 32 MeV/u. Units are relative; darker regions correspond to smaller values of ${\cal E}_2$. \\[-.7cm]}
\end{figure}

\section{Interpreting the results}

Some discussion is now in order for the 50 MeV/u reaction case: while mathematically speaking the absolute minimum corresponds indeed to $\tau=\infty$, one may observe a {\em plateau-like} behavior of this quantity starting at $\tau=50~MeV$. It means
that all values of $\tau$ between 50 MeV and $\infty$ are {\em almost equally} good solutions. 
An illustration of this fact is given in Fig. 6 where charge, number of intermediate mass fragments,
kinetic energy and three highest charged fragment position distributions corresponding to $\tau=$50~MeV and $\infty$ calculated with MMM are compared with the SMF results. The corresponding source parameters are:
(1) $\tau=50~MeV$, $A$=183, $Z$=76, $E$=6.6 MeV/u, $V/V_0$=5.6, $\widetilde{V}/V_0$=6.79, $E_{fl}$=0.6 MeV/u, $\alpha_{fl}$=1.2, $t_p$=58.8 fm/c; (2) $\tau=\infty$, $A$=183, $Z$=76, $E$=6.4 MeV/u, $V/V_0$=9.6, $\widetilde{V}/V_0$=14.78, $E_{fl}$=1 MeV/u, 
$\alpha_{fl}$=1.4, $t_p$=14.8 fm/c. (Here $t_p$ denotes the propagation time, i.e. the time needed to reach,
starting from the configuration predicted by MMM, the {\em dynamical freeze-out} configuration). 
The resemblance between the two cases transpares also from Table I, were the average value of some
observables involved in the construction of $\cal E$, evaluated by MMM for $\tau=50$ MeV, $\infty$ are compared with 
the corresponding SMF ones.
In terms of time along the dynamical evolution, 
the $\tau=50~MeV$ and $\infty$ stages correspond respectively to approximately 140 fm/c and 185 fm/c. Note that for the 50 MeV/u reaction case, the agreement betwen dynamical and statistical radial distributions of the three largest fragments is weaker compared to the case of the 32 MeV/u reaction - see Fig. 7 (Ref. \cite{dyn-1}).

But how to interpret these results? Which is the {\em real} equilibration time? Turning back to Fig. 2 we observe that the major change in ${\cal E}$ versus $\tau$ occurred at $\tau=50~MeV$. 
An interpretation would be that while a major part of the system phase space has already been spanned 
at $t=$140 fm/c (i.e. $\tau=50~MeV$), slight adjustments towards {\em equilibrium} 
are further achieved until $t=$185 fm/c (i.e. $\tau=\infty$).
\begin{table}
\caption{\label{tab:1} MMM: $\tau=50~MeV,\infty$ versus SMF results for various system observables. $\left<{\epsilon} _{IMF}\right>$ is given in MeV/u units.}
\begin{ruledtabular}
\begin{tabular}{cccc}
 Obs. & SMF & MMM: $\tau$=50 MeV & MMM: $\tau=\infty$ \\
$\left<A_{bound}\right>$ & 159.2 & 159.3 & 156.4 \\
$\left<Z_{bound}\right>$ & 66.9 & 68.1 & 67.2 \\
$\left<{\epsilon} _{IMF}\right>$ & 3.65& 3.64& 3.65\\
$\left<Z_{max1}\right>$ & 20.24 & 23.81 & 22.69 \\
$\left<Z_{max2}\right>$ & 13.85 & 14.92 & 15.03 \\
$\left<Z_{max3}\right>$ & 10.62 & 11.21 & 10.35 \\
\end{tabular}
\end{ruledtabular}
\end{table}
\begin{figure}
\includegraphics[height=12cm, angle=0]{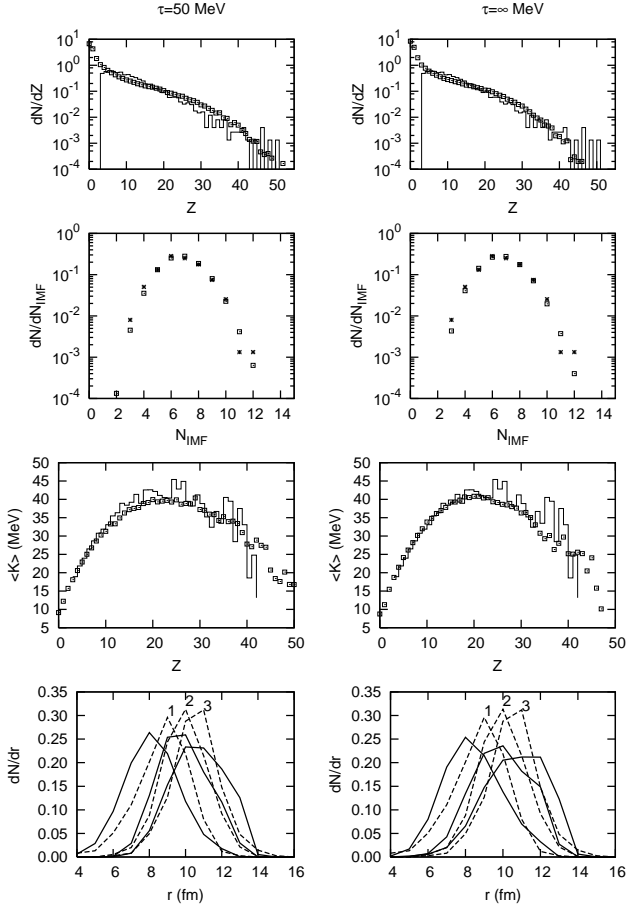}
\caption{From top to bottom: charge, number of intermediate mass fragments (IMF), kinetic, position of the first three highest charged fragment distributions (the position distributions are normalized by $4 \pi r^2$). Left column: $\tau=50~MeV$ MeV; right column: $\tau=\infty$ MeV. MMM results (squares - first three rows from the top; solid line - last row) are plotted in comparison with the SMF ones. Results are corresponding to the $^{129}$Xe+$^{119}$Sn at 50 MeV/u reaction.\\[-.7cm]}
\end{figure}

\begin{figure}
\includegraphics[height=7cm, angle=270]{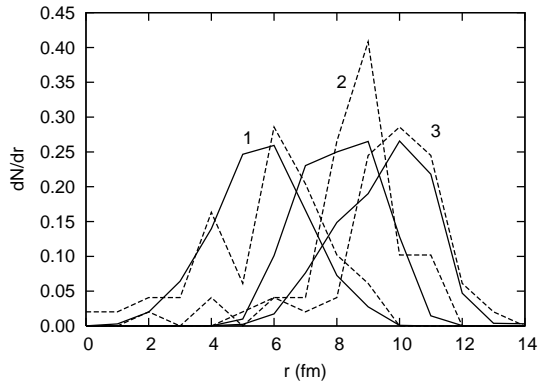}
\caption{Distributions of the position of the first three highest charged fragments (normalized by $4 \pi r^2$). MMM results (solid line) are plotted in comparison with the SMF ones (dashed lines). Results correspond to the $^{129}$Xe+$^{119}$Sn at 32 MeV/u reaction \cite{dyn-1}.\\[-.7cm]}
\end{figure}

\begin{figure}
\includegraphics[height=9cm, angle=270]{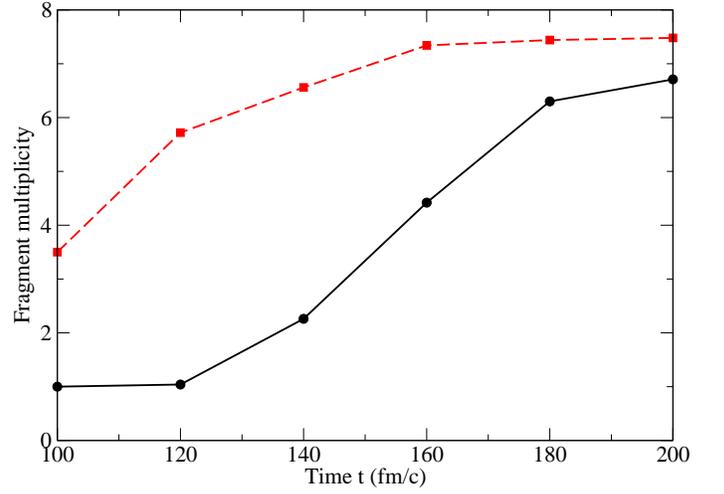}
\caption{(Color online) Dynamical results: Evolution of IMF (circles) and pre-fragment (squares+dashes) multiplicities in time.\\[-.7cm]}
\end{figure}

\begin{figure}
\includegraphics[height=4.5cm, angle=0]{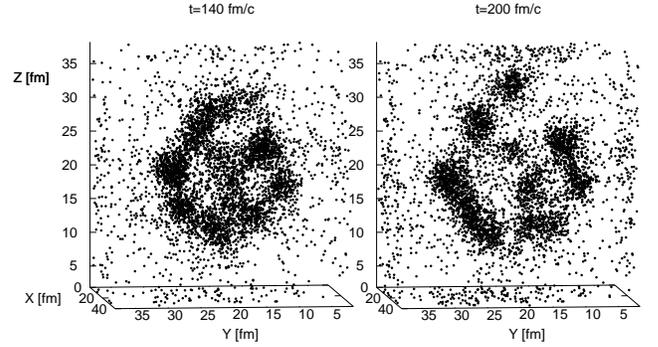}
\caption{Spatial distribution of test-particles in SMF corresponding to the moments of 
time 140 fm/c (left) and 200 fm/c (right) for an arbitray fragmentation event.
\\[-.7cm]}
\end{figure}

Two facts are strongly supporting this hypothesis:
(1) 
As done for the $32~MeV$ case \cite{dyn-1}, one can perform, along the dynamical evolution, 
an analysis in terms of {\em pre-fragments}, i.e. try to identify
density bumps (whose density is larger than the average) that will finally develop into the actual fragments
observed at the $dynamical~ freeze-out$ time \cite{Dorso,AlfioBOB}. 
The SMF number of {\em pre-fragments} saturates at 140 fm/c (see Fig. 8). This is in agreement with Ref. \cite{dyn-1} findings where the equilibration time for the $^{129}$Xe+$^{119}$Sn at 32 MeV/u was found to coincide with the {\em pre-fragment} number saturation point.

(2) While at 140 fm/c the system is still in a pre-fragment stage, and therefore there is strong interaction
between the fragment seeds, at 185 fm/c fragments are almost completely formed and separated. This is clearly illustrated by Fig.9 where fragment spatial configurations yielded by SMF, corresponding to the two moments of time are plotted. From this perspective, the further adjustments towards equilibrium taking place until 185 fm/c are achievable through a strongly decreasing nuclear inter-fragment interaction and Coulomb repulsion.

As a test for the validity of the analysis in terms of pre-fragments, 
we represent in Fig. 10 the distribution of the first three highest charged {\em (pre)fragments} for two moments of time: $t=$140 fm/c and $t=$ 200 fm/c. As it can be observed in Fig. 10 there is a perfect agreement
between the distributions corresponding to the considered moments of time. This fact proves that the
(pre)fragments partitions remain practicaly unchanged during the expansion from the {\em statistical freeze-out} to the {\em dynamical freeze-out}, nicely supporting the working hypothesis of our analysis.

While fairly acceptable, the general agreement between the SMF results and the MMM ones for $^{129}$Xe+$^{119}$Sn at 50 MeV/u reaction is weaker compared to the one corresponding to the $^{129}$Xe+$^{119}$Sn at 32 MeV/u reaction. In the latter case the very good agreement between {\em statistical} and {\em dynamical} results for all the considered observables,
is a strong indication of the system full equilibration.
 In particular, the distribution of the position of the three highest 
charged
fragments, that is reported in Fig.7 for the 32 MeV/u case, is  well reproduced, while the agreement is worse at 50 MeV/u (compare the bottom part of Fig.6 and Fig.7).
Part of this difference is due to the approximations involved in both approaches. In fact, a $20\%$  higher 
flow energy is observed at the {\em dynamical} freeze-out time as compared to the {\em statistical} one, suggesting
some topological differences between the freeze-out partitions of the two approaches.
The other part of it may be simply due to the fact that in the case of the 50 MeV/u reaction the system phase space is only {\em almost} spanned. This could be attributed  
to the fact that the present reaction is more violent compared to  the 32 MeV/u one and, 
therefore, ``more'' dynamically driven. In this respect, see for example the more evident hollow configuration corresponding to 
the dynamical events (Fig.6-bottom panels and Fig.9).

\begin{figure}
\includegraphics[height=10cm, angle=0]{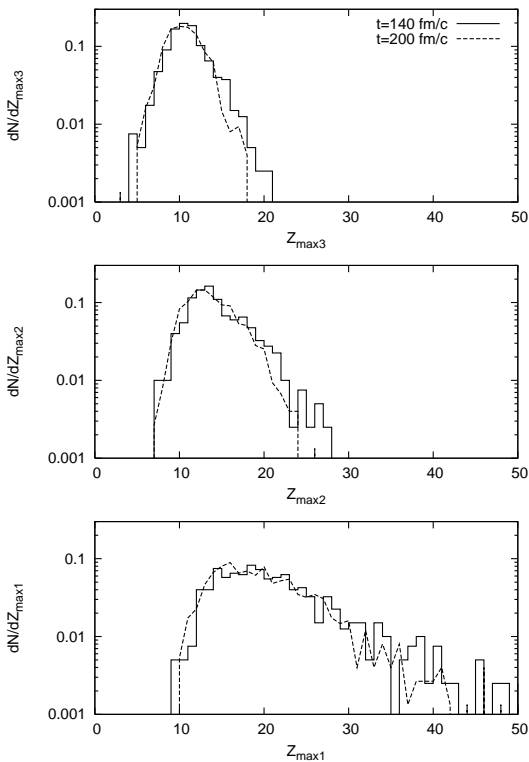}
\caption{Distributions of the first three highest charged {\em pre-fragments} corresponding to two moments of time of the reaction Xe+Sn at 50 MeV/u. Full lines correspond to 
$t=140$ fm/c; dashed lines correspond to $t=200$ fm/c.\\[-.7cm]}
\end{figure}

\section{Final remarks}
Summarizing, the statistical analysis on the SMF dynamical path, started in Ref. \cite{dyn-1}, was extended to the $^{129}$Xe+$^{119}$Sn at 50 MeV/u reaction. While most of the system configuration space appear to be already 
spanned at 140 fm/c, slight adjustments towards equilibration are further achieved until 185 fm/c. 

The following conclusions about the statistical equilibration mechanism in dynamical paths can be drawn: 1) {\em equilibrium
occurs while system constituents are still in interaction};
2) {\em an external radial flow constraint is necessary at 50 MeV/A.}
For both $^{129}$Xe+$^{119}$Sn at 32 MeV/u \cite{dyn-1} and 50 MeV/u reactions (most) of the equilibrium is 
already reached in a {\em pre-fragment} stage
of the system. For both reactions this stage was found to coincide with the saturation of the {\em pre-fragment} number. 
Remarkably, this time is identical for both 32 MeV/u and 50 MeV/u reactions: phase space is spanned   
after the system has spent approximately 
50 fm/c inside the unstable region of the liquid-gas phase diagram and corresponds to about 140 fm/c after the beginning of the reaction. Indeed, this time is related to the time scale of
the instability growth towards the pre-fragment formation \cite{mac,rep}.
However, the corresponding freeze-out volume is  $5.6 V_0$, much larger than the one corresponding to the 32 MeV/u reaction, due to the 
larger compression-expansion effects present in the 50 MeV/u reaction case.  
Further, it should be noticed that, while in the $32$ MeV/u reaction case the equilibrated configuration corresponds to 
$\tau = \infty$ (see Ref. \cite{dyn-1}), here, for the $50$ MeV/u reaction, we find $\tau = 50$ MeV. The physical meaning of this parameter, 
and its relation to the density and temperature conditions reached along the reaction path, 
need to be further investigated. 

The present analysis reveals a qualitative difference between the 32 MeV/u and 50 MeV/u reactions: While in the 32 MeV/u case \cite{dyn-1} fragment partitions 
 and excitation energies impose strong constraints already
identifying the equilibrated source, in the 50 MeV/u case, a subsequent kinematic analysis is necessary. This may be due to the higher amount of {\em liquid}-like phase in the 32 MeV/u reaction (see e.g. the more pronounced shoulder-like shape in the 32 MeV/u charge distribution of 
Ref.\cite{dyn-1}), adding extra fragment-size related constraints to the 32 MeV/u events compared to the 50 MeV/u ones. This detail should be instructive for multifragmentation studies dealing with identifying the equilibrated sources by fitting the data with statistical models.

Finally, the present results may be useful for the experimental nuclear thermodynamics studies  
where volume is a key variable for locating the process in the phase diagram. 

\begin{acknowledgments}
This work was supported by the European Community under a Marie Curie fellowship, contract n. MEIF - CT - 2005 - 010403  
\end{acknowledgments}

\end{document}